\documentclass[12pt]{spieman}  
\usepackage{amsmath,amsfonts,amssymb}
\usepackage{graphicx}
\usepackage{setspace}
\usepackage{tocloft}
\usepackage[flushleft]{threeparttable}

\title{On the Feasibility of Using a Laser Guide Star Adaptive Optics System in the Daytime}

\author[a]{Ryan Dungee}
\author[b]{Mark Chun}
\author[c]{Yutaka Hayano}
\affil[a]{University of Hawaii, Institute for Astronomy, 2680 Woodlawn Drive, Honolulu, United States, 96822}
\affil[b]{University of Hawaii, Institute for Astronomy, 640 North A`oh\={o}k\={u} Place, Hilo, United States, 96720}
\affil[c]{National Astronomical Observatory of Japan, Mitaka, Tokyo, 181-8588, Japan}

\cftpagenumbersoff{figure}
\cftpagenumbersoff{table}
\begin{document} 
\maketitle

\begin{abstract}
We investigate the use of ultra-narrow band interference filters to enable daytime use of sodium laser guide star adaptive optics systems. Filter performance is explored using theoretical and vendor supplied filter transmission profiles, a modeled daylight sky background, broadband measurements of the daytime sky brightness on Maunakea, and an assumed photon return from the sodium laser guide star and read noise for the wavefront sensor detector. The critical parameters are the bandpass of the filter, the out-of-band rejection, and the peak throughput at the wavelength of the laser guide star light. Importantly, a systematic trade between these parameters leads to potentially simple solutions enabling daytime observations. Finally, we simulated the Mid-Infrared Adaptive Optics (MIRAO) system planned for the Thirty Meter Telescope with an end-to-end simulation, folding in daytime sky counts. We find that MIRAO with five sodium laser guide stars, commercial off-the-shelf filters to suppress the sky background in the laser guide star wavefront sensors, and a near-infrared natural guide star ($K \le 13$) tip/tilt/focus wavefront sensor can attain daytime Strehl ratio values comparable to those at night.
\end{abstract}


{\noindent \footnotesize\textbf{*}Ryan Dungee,  \linkable{rdungee@hawaii.edu} }

\begin{spacing}{2}   

\section{Introduction}
\label{sec:intro}

At wavelengths longward of $\sim\!2\;\mu\mathrm{m}$, the sky background in ground-based observations begins to be dominated by the thermal emission of the telescope and the atmosphere\cite{b&c}. Around $\sim\!3\;\mu\mathrm{m}$, this thermal background is so strong that the sky background does not change appreciably between the night and the day. As a result, ground-based observatories can perform mid-infrared (mid-IR) observations during the day with little impact on the sensitivity of the observation\cite{b&c}. Some facilities, such as the NASA Infrared Telescope Facility (IRTF), routinely make observations during the day as a way of extending the science time available to their users. This practice is not common as there are a few extra complications associated with operating during the day. Nonetheless, early morning hours which are no longer suitable for optical observations can be used for IR observing to extend the operation time of a telescope with little to no additional engineering.  Even an hour of additional time during the morning twilight would increase the total telescope time available for science observations by $\sim 10\%$. Given the construction and operating costs of the current and next generation of large ground-based telescopes, such an increase in time is extremely valuable.

However, for the next generation of extremely large telescopes there is a secondary consideration since they require adaptive optics (AO) systems at mid-IR wavelengths to reach the diffraction limit\cite{b&c}. The Thirty Meter Telescope (TMT), one of the next generation ground-based telescopes, is being designed with AO as an integral part of the facility. The first-light AO system (NFIRAOS)\cite{nfiraos1,nfiraos2}, optimized for observations at near-IR wavelengths, will use a sodium laser guide star (LGS) system in order to maximize the area of sky where they can obtain diffraction-limited observations.  The wavelength of the $\mathrm{D_2}$ transition in sodium, $589.0\;\mathrm{nm}$, falls well within the visible range meaning the return signal from the laser guide star would normally be drowned out by the daytime sky. However, the spectral bandwidth of the sodium LGS return light is on the order of a few picometers, so in principle a sufficiently narrow spectral bandpass filter can be used to isolate the LGS light from the sky background.

In 2001, Beckers and Cacciani proposed the use of a sodium LGS system during the daytime with applications in both solar astronomy and in the mid-IR\cite{b&c}. Beckers revisited these ideas in 2008 laying out both site specific considerations and the use of such mid-IR instruments in the era of James Webb Space Telescope\cite{jusb}. In 2016, Hart et al. performed the first measurement of the achievable contrast between an LGS and the daytime sky\cite{hart}. For this experiment Hart et al. used a magneto-optical filter (MOF), as suggested by Beckers and Cacciani. An MOF uses sodium vapor surrounded by a strong magnetic field to rotate the polarization of light resonant with transitions in the sodium. By using crossed linear polarizers at the front and rear of the sodium cell and a vapor temperature/magnetic field configuration that rotates the polarization of the resonant light by $90^\circ$. All wavelengths not interacting with the magnetized sodium atoms are attenuated at the level of the crossed polarizers ($\sim\!10^6$). This allows the filter to attain extremely narrow bandwidths.

However, the MOF approach carries with it a number of disadvantages: (1) it is a large bulky device, (2) it requires a heated sodium cell, and (3) the overall throughput is low.  The use of a such a filter within a wavefront sensor will require substantial engineering to fit it into the wavefront sensor and to mitigate any localized turbulence created by the sodium vapor which has temperatures ranging from $150$ to $200^{\circ}\;\mathrm{C}$.

In order to address these disadvantages we explored (1) the trade-offs between different sodium line filter specifications and their impact on the signal-to-noise ratio (SNR) of the wavefront sensor measurements and (2) how this SNR translates into AO performance using detailed simulations of the full system. In the rest of this paper we discuss the results of our filter specification trade study (Sec. \ref{sec:filter-req}), the AO performance simulations (Sec. \ref{sec:simu}), discuss particular concerns and implications of the proposed approach (Sec. \ref{sec:disc}), and present our conclusions on the feasibility of daytime LGS AO (Sec. \ref{sec:conc}).

\section{Filter Requirements}
\label{sec:filter-req}

There are many sources of error that impact the performance of an AO system. For this work the pertinent sources of error are encapsulated in the wavefront sensor (WFS) measurement error.  Here we consider a Shack-Hartmann WFS (SHWFS) as it is the most common type of wavefront sensor in use with sodium LGSs.  In a SHWFS the wavefront gradient measurement error is proportional to the surface brightness and subaperture spot size, and inversely proportional to the SNR\cite{hardy}. In this work, we make the assumption that the WFS spot size is independent of the approach we take to filtering the sky background and as such we focus on the changes to SNR. The SNR depends on the laser return rate, effective collecting area of each subaperture, sky background, and sensor readout noise.  As a first step we explored how basic filter parameters affect the SNR. These SNR values are calculated using a standard formula, where the terms and their assumed values are defined in Table \ref{tab:eq-parts}:

\begin{table}
\caption{Symbols for Equation 1}
\label{tab:eq-parts}
\centering

\begin{threeparttable}
\begin{tabular}{|r|c|c|c|}
\hline
\rule[-1ex]{0pt}{3.5ex} Variable & Description & Assumed Value & Value Reference \\
\hline
\hline
\rule[-1ex]{0pt}{3.5ex} S & Signal & $900\tnote{a} \times A\tnote{b} \times T_{589}\tnote{c}$ & Boyer et al.\cite{lgsdoc} \\
\rule[-1ex]{0pt}{3.5ex} B & Background & Detected photons from the sky & Our calculations \\
\rule[-1ex]{0pt}{3.5ex} $n_{pix}$ & Pixels per subaperture & $6 \times 16$ pixels\tnote{d} & Ellerbroek et al.\cite{TMTAO} \\
\rule[-1ex]{0pt}{3.5ex} RN & RMS detector Read Noise & $3\;\mathrm{e^-}$ & Ellerbroek et al.\cite{TMTAO} \\
\hline
\end{tabular}

\begin{tablenotes}
\item[a] Photo-detection events per subaperture per frame in the wavefront sensor for the Narrow Field Infrared Adaptive Optics System (NFIRAOS).
\item[b] The collecting area of a subaperture is taken to be an adjustable parameter and is dependent on the order of the system.  Note that here it is expressed as an area relative to the NFIRAOS subaperture size (e.g. $A$ of 1 means a subaperture size of $0.25\;\mathrm{m}^2$).  
\item[c] $T_{589}$ is the transmission of a given filter at $589.0\;\mathrm{nm}$, the specified return wavelength of the laser guide star (LGS).
\item[d] We always assume the largest detector size to make our estimates conservative.
\end{tablenotes}

\end{threeparttable}
\end{table}

\begin{equation}
\label{eq:snr}
SNR = \frac{S}{\sqrt{S+B+n_{pix} RN^2}}
\end{equation}

The signal $S$ is derived from the TMT specification for the sodium laser guide star return for NFIRAOS and corresponds to an LGS magnitude of $R\approx8.2$, a value routinely achieved by Keck II\cite{keckLGS}. The number of photons received from the LGS is related to the subaperture collecting area, $A$, which also determines the correction order of the AO system. Thus, increasing the size of the subapertures provides for a higher LGS signal, increasing the SNR but reducing the order of the AO correction. Since this work considers the overall feasibility, we leave the detailed optimization of this trade to another study.  The peak transmission of the filter at $589.0\;\mathrm{nm}$, $T_{589}$, accounts for the reduction in the transmission by the inclusion of our filter. The background $B$ is calculated from our model sky spectrum convolved with our selected filter transmission profile. Finally, the read noise, $RN$, is derived from the TMT specification for the read noise per pixel for the NFIRAOS wavefront sensor cameras and an assumption of the number of pixels covering the subaperture field of view. This sets the read noise at 3 electrons per pixel when operating at a frame rate of $800\;\mathrm{Hz}$. Detector sizes range from $6\times 6$ pixels to $6\times 16$ pixels depending on the LGS's elongation in the subaperture's field of view. Recent tests of detectors for the TMT wavefront sensors have achieved a read noise between 3 and 3.5 electrons\cite{TMTAO}. However, we note that given the bright, daytime sky background, we are not read noise limited and, as our later discussions will demonstrate (Sec. \ref{sec:simu}), the achieved range of read noise values are well within an acceptable range. Our model for the sky background will be discussed further in Sec. \ref{subsec:onsky}.  

\subsection{On-Sky Tests}
\label{subsec:onsky}

\begin{figure}
  \centering
  \includegraphics[width=3.3125in]{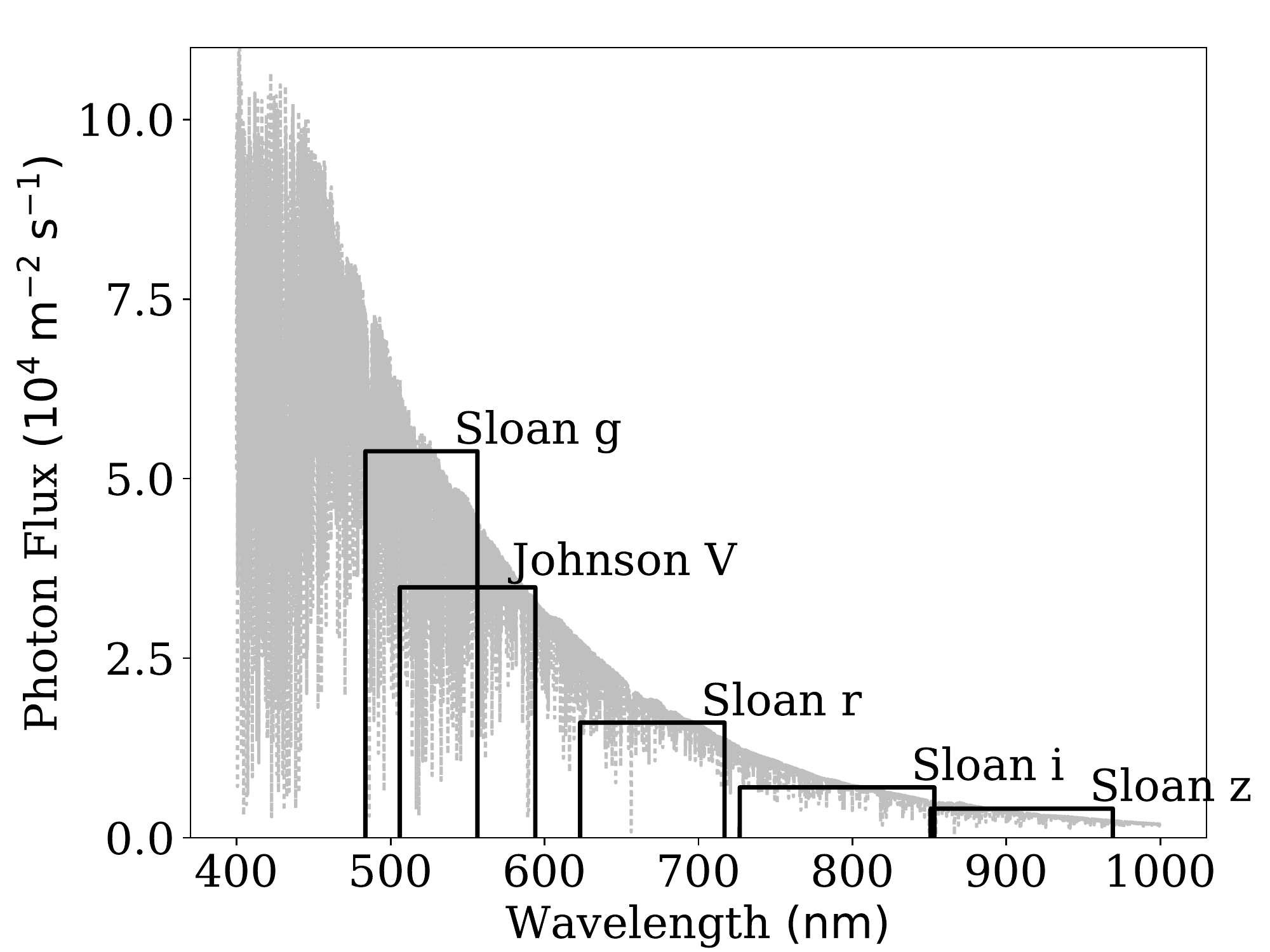}
  \caption{ \label{fig:modelspectrum} 
In gray, the model spectrum we use as input for our SNR calculations. Solid black lines plotted over the spectrum are our measurements (see Table \ref{tab:skymags}), converted from magnitudes into the number of photons our wavefront sensor would measure through that filter's passband and binned at the same resolution as the spectrum. For further details, see Sec. \ref{subsec:onsky}. }
\end{figure}

\begin{table}
\caption{Measured Sky Brightnesses}
\label{tab:skymags}
\centering

\begin{threeparttable}
\begin{tabular}{|r|c|}
\hline
\rule[-1ex]{0pt}{3.5ex} Filter Name & Magnitude per Square Arcsecond \\
\hline
\hline
\rule[-1ex]{0pt}{3.5ex}Sloan g & $3.4\pm0.1$ \\
\rule[-1ex]{0pt}{3.5ex} r & $4.5\pm0.1$ \\
\rule[-1ex]{0pt}{3.5ex} i & $5.2\pm0.1$ \\
\rule[-1ex]{0pt}{3.5ex} z & $5.6\pm0.1$ \\
\rule[-1ex]{0pt}{3.5ex} Johnson V & $3.9\pm0.1$ \\
\rule[-1ex]{0pt}{3.5ex} MKO K & $7.2\pm0.1$ \\
\hline
\end{tabular}
\end{threeparttable}

\end{table}

Our model sky spectrum is derived from broadband measurements of the daytime sky and a solar spectrum provided by the Harvard-Smithsonian Center for Astrophysics\cite{solarspec} modified to have a Rayleigh scattering wavelength dependence of $\lambda^{-4}$. In the visible band we obtained measurements for Johnson V, and Sloan g, r, i, and z broadband filters from the MORIS instrument on IRTF\cite{morispaper}. Photometric calibrations for the instrument were obtained from previous observations of a standard star. Observations were of a patch of clear sky $90^\circ$ away from the sun approximately 3 hours after sunrise. The results of these observations are summarized in Table~\ref{tab:skymags}.

We fit the model sky spectrum to the five broadband measurements using the normalization as our only free parameter. The results (see Fig.~\ref{fig:modelspectrum}) were checked by comparing broadband counts in our model spectrum to their corresponding measurement, indicating our model spectrum is accurate to $\pm20\%$. Overall, our model sky spectrum is conservative with respect to our measurements. Our model's overestimation is readily visualized in Fig.~\ref{fig:modelspectrum}, where each measured sky magnitude was converted into an average photon flux for the corresponding passband. This model sky spectrum is used in our sensitivity study (Sec.~\ref{subsec:sensitivity}) and our simulations (Sec.~\ref{sec:simu}).

\subsection{Sensitivity to Filter Parameters}
\label{subsec:sensitivity}

\begin{figure}
  \centering
  \includegraphics[width=6.75in]{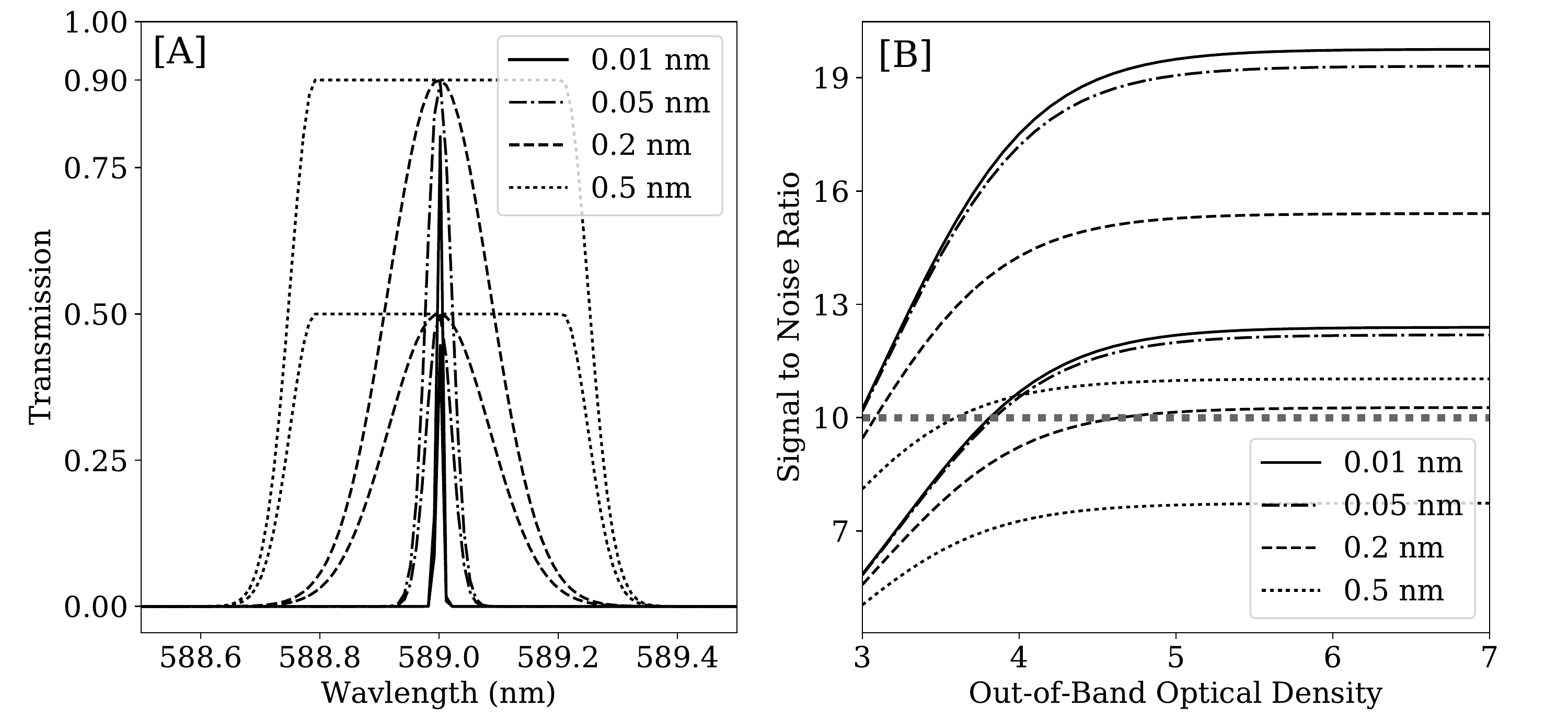}
  \caption{ \label{fig:sens_study}
\textbf{[A]:} examples of the theoretical transmission profiles we used in our sensitivity study, including peak transmissions of 0.5 and 0.9. \textbf{[B]:} The out-of-band rejection of the filter expressed as an optical depth versus the signal-to-noise ratio (SNR). The horizontal dotted line represents our initial cut-off SNR of 10. The AO configuration corresponds to that which is outlined in Table \ref{tab:eq-parts}, with $A$ of 1. } 
\end{figure} 

To test the importance of various filter parameters we performed a sensitivity study by generating a number of theoretical transmission profiles and convolving them with our model daytime sky spectrum. Filter traces were generated with different bandwidths, out-of-band rejections, and peak transmission values. Mock filter traces with a bandwidth less than $0.2\;\mathrm{nm}$ had a Gaussian profile. Whereas filters wider than $0.2\;\mathrm{nm}$ had a roughly box profile, with edges that decayed exponentially. The photon count left after convolution of a profile with our model sky spectrum was then used as the background in Eq.~\ref{eq:snr}. The $T_{589}$ value folded into the signal in Eq.~\ref{eq:snr} was determined by the mock filter's peak transmission. All profiles were assumed to be centered at $589.0\;\mathrm{nm}$, and had a minimum transmission specified by the out-of-band rejection given as an optical density (OD).

In varying the filter parameters, we calculated the SNR of a wide variety of filter transmission profiles. A plot of the important parameters is shown in Fig.~\ref{fig:sens_study}. For our initial performance cut-off of SNR greater than $10$, we find that the bandwidth of the filter must be sub-nanometer to sufficiently block out the sky background but that for bandwidths less than $50\;\mathrm{pm}$ the SNR is photon noise limited. Similarly, out of band rejections greater than OD$\sim\!4.5$ were found to have diminishing returns. The peak transmission is a critical parameter and it allows the bandwidth of the filter to increase. This is illustrated in Fig.~\ref{fig:sens_study} where the $0.5\;\mathrm{nm}$ bandpass filter with a high transmission performs nearly as well as as the $10-50\;\mathrm{pm}$ filters with lower transmission.  Conversely a low peak transmission can be compensated by a smaller filter bandpass. However a peak filter transmission less than about 50\% has difficulty achieving the SNR cut-off.

\subsection{Application to Commercial Filters}
\label{subsec:apptocots}
We obtained transmission profiles for several off-the-shelf and custom filters. Putting these profiles through the same code used in our sensitivity study allowed us to rank their expected performance. The typical resolution provided on a filter's profile was of order $0.06\;\mathrm{nm}$, so to match the resolution of the profiles to our sky spectrum we used linear interpolation on the provided filter profiles.

\begin{table}
\caption{Comparison of various filters}
\label{tab:combined}
\centering

\begin{threeparttable}
\begin{tabular}{|r|c|c|c|c|}
\hline
\rule[-1ex]{0pt}{3.5ex} Filter Name & Expected SNR & Bandwidth & Peak Transmission & Out of band \\
\hline
\hline
\rule[-1ex]{0pt}{3.5ex} Alluxa OD4\tnote{a} & $21.0$ & $0.5\;\mathrm{nm}$ & $>90\%$ & $>\mathrm{OD}\;4$ \\
\rule[-1ex]{0pt}{3.5ex} Custom Filter A & $19.9$ & $\sim\!1.0\;\mathrm{nm}$ & $92.4\%$ & $>\mathrm{OD}\;5$ \\
\rule[-1ex]{0pt}{3.5ex} Alluxa OD6\tnote{b} & $16.2$ & $0.8\;\mathrm{nm}$ & $96.9\%$ & $>\mathrm{OD}\;5.7$ \\
\rule[-1ex]{0pt}{3.5ex} Keck Sodium Notch Filter & $13.8$ & $1.0\;\mathrm{nm}$ & $90.9\%$ & $>\mathrm{OD}\;4$ \\
\rule[-1ex]{0pt}{3.5ex} MOF\tnote{c} & $10.8$ & $\sim10\;\mathrm{pm}$ & $<50\%$\tnote{d} & $\sim\!\mathrm{OD}\; 6$\tnote{e} \\
\hline
\hline
\end{tabular}

\begin{tablenotes}
\item[a] The filter's full name is Alluxa 589.16 OD4 Ultra Narrow
\item[b] The filter's full name is Alluxa 589.45 OD6 Ultra Narrow
\item[c] Magneto-optical filter (MOF), these are estimates based off reported performance from outside sources\cite{hart} and private discussions with Wayne Rodgers of Eddy Co. a maker of MOFs.
\item[d] Return light of the laser is circularly polarized but the MOF construction requires linear polarizers. This value can be boosted by the inclusion of a quarter wave-plate\cite{hart}.
\item[e] This depends on the polarizers chosen for the MOF construction.
\end{tablenotes}

\end{threeparttable}
\end{table}

The results of these calculations are shown in Table \ref{tab:combined}. 
Included in the table are the characteristics of the MOF filter for comparison. Interference filters have the advantage of being a passive optical element with a relatively simple implementation in the AO system. They also typically have much higher peak transmission values. Their principal disadvantage lies in their dependence on the angle of incidence of the beam on the filter which can lead to a relatively narrow field of view.

Of the interference filters analyzed, the Alluxa 589.16 OD4 Ultra Narrow Bandpass Filter (or `Alluxa OD4') achieved the highest SNR and became the focus of further simulations. As such, much of the following discussion is built around this particular filter's expected performance. Nonetheless, the SNR is a generic enough measure that these discussions can be expected to apply to other filters capable of matching the Alluxa OD4's performance.

\begin{figure}
  \centering
  \includegraphics[width=3.3125in]{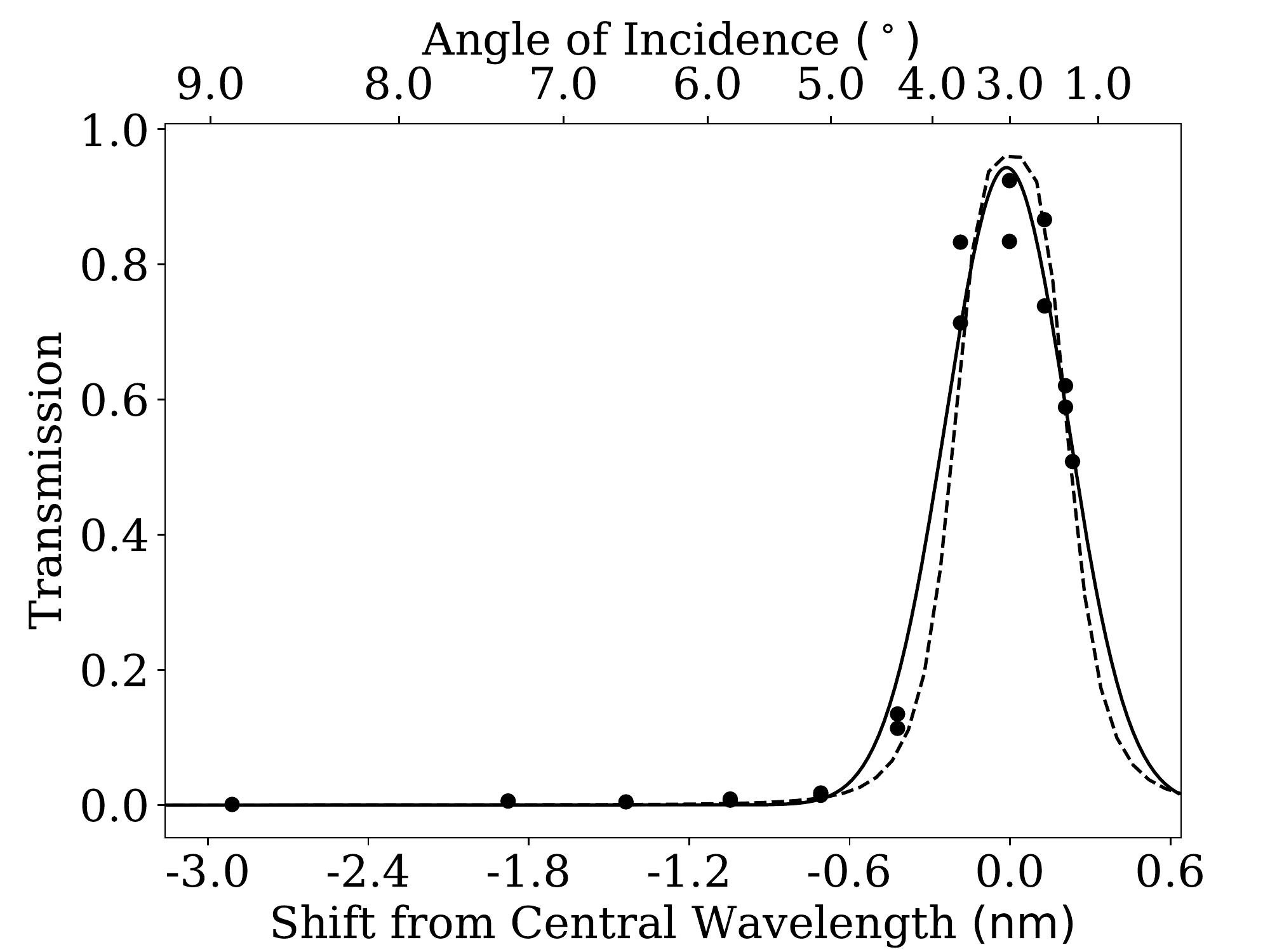}
  \caption{ \label{fig:yutakacurve} 
Filter profiles for the Alluxa OD4 filter. Black dots are measurements, taken in steps of $1^\circ$ by tilting the filter with respect to the laser beam which is at a the wavelength of the sodium $\mathrm{D}_2$ line. The solid black curve is the best fit filter profile assuming a Gaussian curve. The dashed black line is the filter profile provided by Alluxa. Shifts from the central wavelength are computed using the standard approximation $\lambda(\theta)=\lambda_0\;\sqrt[]{\frac{1-x(\theta)}{1-x(-3^\circ)}}$. Where $x(\theta) = \frac{\sin \theta}{n_{eff}}$ with $n_{eff}=1.85$ is the effective refractive index for the filter's material and $-3^\circ$ is the angle at which transmission through the filter peaked\cite{alluxan}. }
\end{figure}

The filter profile and field of view for the Alluxa OD4 were measured on the optical bench of the Subaru telescope's sodium laser. Our measurements show that the filter profile is wider than reported by the manufacturer by a factor of $\sim\!1.5$ the full width at half maximum (FWHM). We have yet to measure the out of band rejection but preliminary results from measuring the profile indicate it reaches the specifications of OD $>$ 4. To measure the filter's profile we measured throughput as a function of the angle of incidence on the filter, and fit the data points with a Gaussian curve to obtain the FWHM, peak transmission, and central wavelength. This yields an estimated FWHM $0.76\;\mathrm{nm}$ versus the provided value of $0.5\;\mathrm{nm}$, an estimated peak transmission of $0.94$ versus the provided value of $0.96$, and an estimated central wavelength of $589.14\;\mathrm{nm}$ that agrees with the provided value. For a comparison of the measured and provided filter profiles we refer the reader to Fig.~\ref{fig:yutakacurve}.

As with any interference filter, the central wavelength (and by extension the throughput at the wavelength of the sodium $D_2$ line) varies as a function of the angle of incidence on the filter. This variation with field angle will come into play for WFSs with a small beam size (e.g.~large pupil demagnifications). Our measured filter profile shows a throughput greater than 85\% for collimated beams with angles of incidence $\pm 0.7^\circ$ from normal. If we first consider a 10-meter telescope with a collimated optical beam size of $10\;\mathrm{mm}$ (e.g. a standard one-inch filter) and a subaperture field of view of 2 arcseconds, then we have a requirement of $\pm 0.3^\circ$.  This field propagates through the Alluxa OD4 with little or no field attenuation.  However, for next generation large ground-based telescopes such as TMT, the limitation on the field of view will need to be considered due to further elongation of the LGS by the larger aperture and a greater pupil demagnification. We discuss options to address this in Sec.~\ref{disc:truncation}.

\section{Simulations}
\label{sec:simu}

Since the SNR is not a direct measurement of the AO system's performance we also ran end-to-end simulations of the TMT MIRAO system. These simulations include the effect of daytime sky on the sodium LGS wavefront sensors as well as the effect of the daytime sky on the low-order (tip/tilt/focus) natural guide star (NGS) wavefront sensor that is required with an LGS AO system.  They do not include the variation of the filter transmission as a function of field angle. We leave this to a future study.

Simulations are performed using the Multi-threaded Adaptive Optics Simulator (MAOS) package\cite{maos}. MAOS is an open source Monte-Carlo adaptive optics simulation package created and maintained by the TMT Corporation. The package includes simulation of the wavefront propagating through the atmosphere, the correction by the deformable mirror, the wavefront sensing by a variety of LGS and NGS wavefront sensors, and the algorithm to reconstruct the wavefront.  We add in additional background counts to account for the added sky brightness during the daytime. Since the NGS in general has no sufficiently bright spectral feature the WFS must use a broadband filter preventing a daytime sky suppression analogous to our LGS component. Because of this we benefit from working at longer wavelengths (e.g. K-band) where the image is partially corrected by the AO system. The image will then have a well corrected diffraction-limited core ($\sim15\;\mathrm{mas}$) allowing us to, in principle, suppress the sky background through a careful selection of the pixel scale and field stop size in the WFS. For this feasibility study, we did not optimize the number of subapertures but rather we sought to demonstrate that there is a stable and useful configuration that enables daytime observing.

\subsection{Simulation Setup}

For our simulations, the sky backgrounds are calculated and added as a uniform background of photo-detection events per pixel per frame with Poisson noise. Atmospheric conditions are modeled as a set of evolving phase screens which each add aberrations to the simulated wavefront as it passes through the layer. Each phase screen has a random set of aberrations with a Von Karman spectrum and a wavefront outerscale of 30 meters. Given that there is some randomness to each realization we average the output of 20 independent runs of the simulation. The difference between each independent run is the set of atmospheric screens used. The same 20 sets of atmospheric screens were used for every AO and filter configuration. This allows us to account for variations of the seeing in each realization due to the limited number of simulation steps. We estimate the uncertainty in the resulting mean wavefront error is $(\phi^2)^{+0.02}_{-0.03}\;\mathrm{rad}^2$  where $\phi^2$ is the phase variance of the wavefront, or roughly an uncertainty of $\pm0.01$ on the Strehl ratios that we calculated. We determine this estimate from the variance of the mean wavefront error as a function of the number of realizations binned together compared to the averaged output of a set of 70 simulation runs. For characterizing the daytime AO system, we ran batches of simulations for two separate atmospheric conditions, the 50th and 75th percentile seeing cases for Maunakea at 13N. These conditions are based on the TMT site testing campaign\cite{schoeck} and represent median conditions and worse than average night time conditions.  We note that we expect daytime observing with sodium LGSs on TMT would be limited to the early morning.  During this time there is evidence that the seeing is very similar to the end of the night values for about one-two hours\cite{early-seeing}.

\subsection{Stable AO Performance}
\label{subsec:high-order}

\begin{figure}
  \centering
  \includegraphics[width=3.3125in]{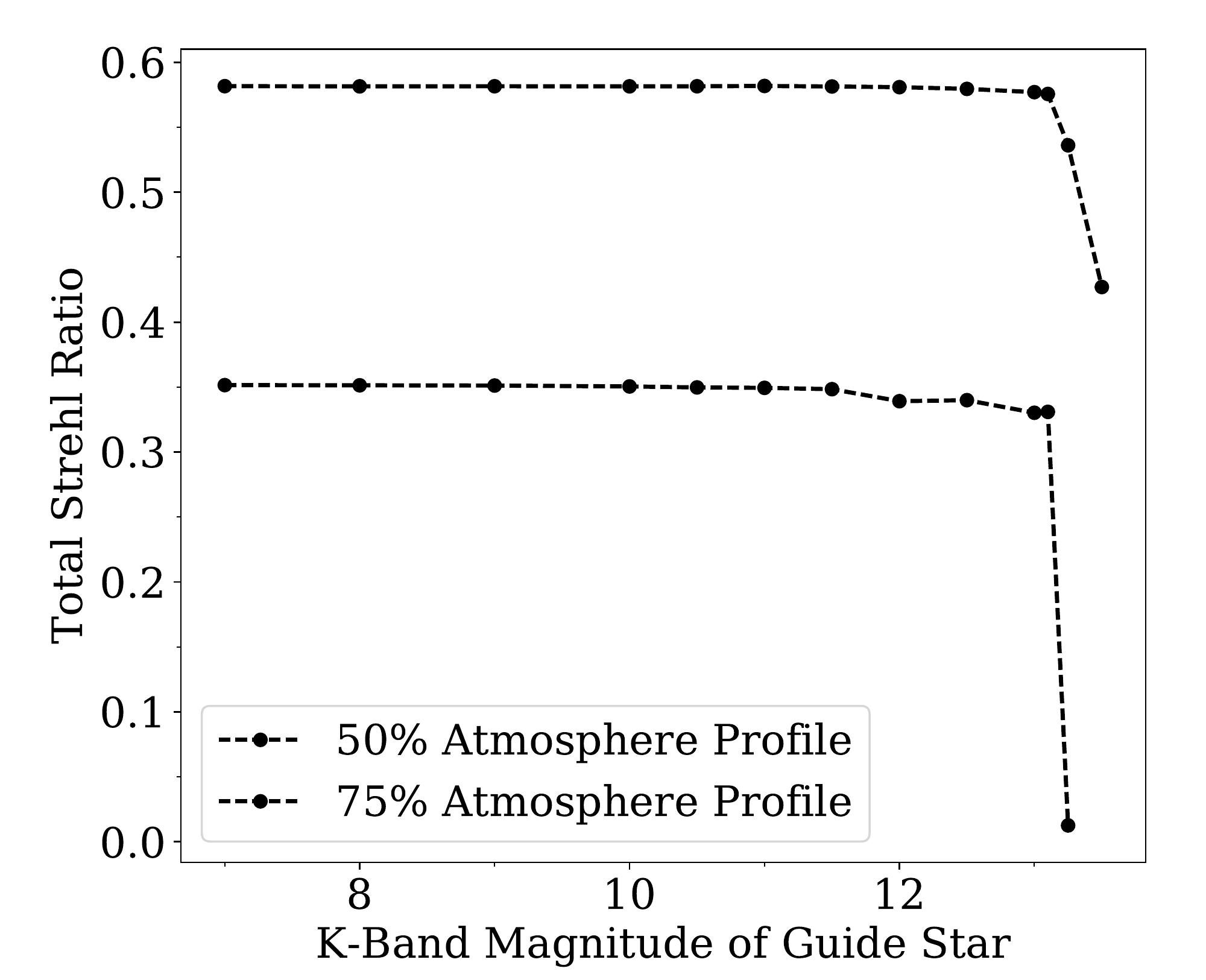}
  \caption{ \label{fig:totalstrehl}
The total K-band Strehl ratio versus magnitude of the NGS for the MIRAO system (see Sec. \ref{subsec:high-order}) in daytime conditions. The sky background for the NGS (tip,tilt,focus) wavefront sensor corresponds to our measured MKO K magnitude value. The sky background for the LGS wavefront sensor corresponds to the expected flux through the Alluxa OD4 filter. The sharp decrease in performance at the $K\sim 13$ results from the rapid decrease in the SNR of the NGS wavefront sensor against the day time sky. }
\end{figure}

We split the performance calculations into two steps:  identifying a configuration of the high-order LGS wavefront sensors that provided good correction in the mid-IR during the day, and the determination of a limiting magnitude for the NGS wavefront sensor which measures tip/tilt/focus modes that can not be obtained from the LGSs.  We based our simulations on MIRAO, an AO system that aims to have low thermal emissivity while providing corrections in the mid-IR wavelengths.  The current design calls for three LGSs and a single NGS used for tip/tilt/focus sensing.  The system has a $60\times60$ subaperture array with a single deformable mirror operating at $800\;\mathrm{Hz}$.  

With a daytime sky background the SNR in MIRAO's wavefront sensors proved too low for stable operation of the AO system. As such, a new configuration was chosen to boost the SNR for the LGS wavefront sensor with the Alluxa OD4 filter to a point we were certain the AO system would be stable. We chose a SNR value of $20$ with the expectation that it would be robust to variations in the sky background.  This corresponded to a factor of $3$ in subaperture area resulting in a $34\times34$ subaperture array. Additionally, two LGSs were added to the system in order to reduce the LGS focal-anisoplanatism in order to boost the performance at shorter wavelengths (e.g.~K-band) where the NGS wavefront sensor was chosen to operate. For all of the simulations presented here we assume that the science object is also used as the NGS. These changes led to stable operation of the system, and so the final configuration is a 5 LGS, 1 NGS system with $34\times34$ subapertures running at $800\;\mathrm{Hz}$ on a single deformable mirror.

For the NGS wavefront sensor we chose $K$-band (i.e.~2.2 microns) to benefit from the AO corrected point spread function (PSF) and from a lower sky background.  However, we did not perform an optimization here and shorter or longer wavelengths could be considered.  Our simulations used K-band sky brightness values measured using the near-IR guider camera of the iShell instrument on IRTF. Observations of the star HD377 were made in clear conditions. This was done at $07\!:\!42\;\mathrm{AM}$ local time with the star approximately $90^\circ$ away from the Sun. The result of these measurements can be found in Table \ref{tab:skymags} along with the other sky brightness measurements.

\begin{table}
\caption{Strehl Ratio at Science Wavelengths of Interest}
\label{tab:sr-ratios}
\centering

\begin{threeparttable}
\begin{tabular}{|r|c|}
\hline
Wavelength & Total Strehl Ratio \\
\hline
\hline
$2.2\;\mathrm{\mu m}$ & $58\%$ \\
$3.8\;\mathrm{\mu m}$ & $83\%$ \\
$4.7\;\mathrm{\mu m}$ & $89\%$ \\
$10.8\;\mathrm{\mu m}$ & $98\%$ \\
$13.0\;\mathrm{\mu m}$ & $99\%$ \\
\hline
\end{tabular}
\end{threeparttable}
\end{table}

The performance of the AO system demonstrated in Fig.~\ref{fig:totalstrehl} reflects that of a robust (see Sec.~\ref{disc:skyvariations}) daytime system which achieves near optimal (i.e. nighttime) performance down to NGS magnitudes of $K\sim\!13$. Table \ref{tab:sr-ratios} contains a list of attainable Strehl ratios at longer wavelengths. For brighter NGS magnitudes, the performance is driven by the LGS component of the system since the reduction in the Strehl ratio from the corrected low order aberrations is less than $1\%$. For fainter natural guide stars there is a sharp drop in performance.  This sharp drop is partially from the fact that the NGS wavefront sensor parameters were not optimized for fainter guide star magnitudes. However, even with an optimization we expect the limiting magnitude cut-off to be sharper than that for a nighttime system. To see why, consider a nighttime Shack-Hartmann WFS for which the sensitivity is often limited by read noise so the SNR can improve proportionally with increasing the WFS integration time (SNR $\sim t$).  However, for the daytime case the system is background limited so the SNR increases only as $\sim\sqrt{t}$. Since this study is concerned with demonstrating the feasibility, system parameters were not optimized as a function of guide star brightness. Doing so should extend the limiting magnitude of the system fainter. 

\section{Discussion}
\label{sec:disc}

Our calculations indicate that a relatively simple, inexpensive, interference filter can enable daytime AO with a sodium LGS. In this section we comment on the sensitivity of our approach to variations in the sky brightness, and discuss some important design constraints imposed by the interference filter approach for LGSs.   

\subsection{Sky Brightness Variations}
\label{disc:skyvariations}

\begin{figure}
  \centering
  \includegraphics[width=6.75in]{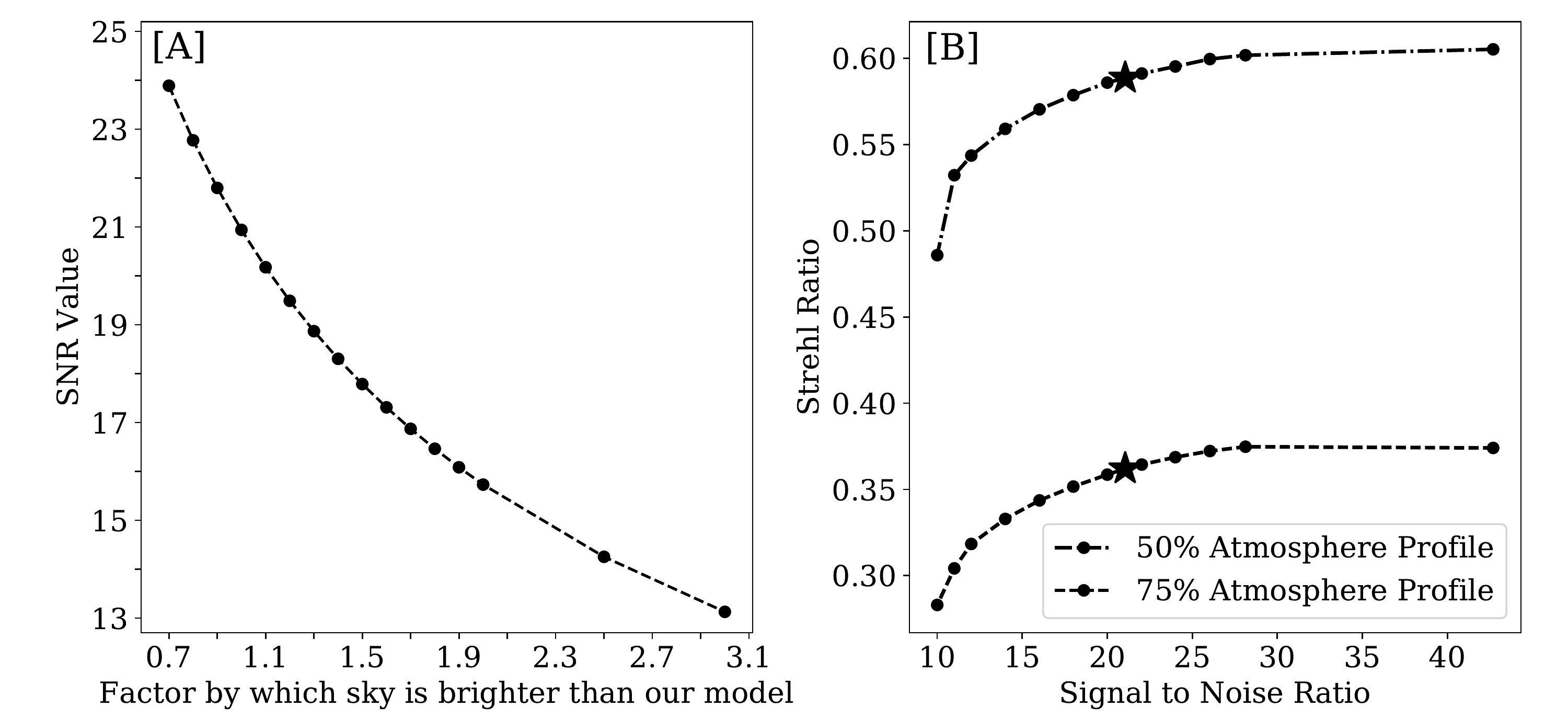}
  \caption{ \label{fig:skysens}
\textbf{[A]}: A plot of the expected signal-to-noise ratio (SNR) of the Alluxa OD4 filter that was chosen in Sec.~\ref{sec:filter-req} versus the sky brightness. The sky brightness is expressed as a multiplicative factor relative to our measured value. For example, a factor of 2 represents 2 times as many background photons measured by the WFS.
\textbf{[B]}: The high order K-band Strehl ratios for the MIRAO system in both the 50th and 75th percentile atmosphere cases (median and poor seeing respectively) versus SNR achieved in the WFS. The SNR is calculated according to Eq.~\ref{eq:snr} with $T_{589}=0.9$ and $A=3$. The star on each curve reflects the expected performance of the Alluxa OD4 filter (SNR of 21). }
\end{figure}

Throughout this paper we have assumed a sky brightness based on a single measurement taken with an observing angle $90^\circ$ away from the Sun which corresponds to the darkest part of the sky. A major objection to this approach would be that the sky brightness will vary with observing angles relative the Sun (look angles), with time of day, with day-to-day variations in aerosol content of the atmosphere, and over the course of a year. We could not find a long-term survey of the daytime sky brightness in the near-infrared or visible bands. However, models of atmospheric scattering processes have been used to explore how near-IR sky brightness varies with time of day, look angle, and by time of year for Haleakala\cite{hale-sim}. Moreover, observations for this site have since proven consistent with these predictions\cite{hale-obs}. If we apply the expected restrictions for a large night-time astronomical telescope, limiting observations to the first few hours of the morning and to look angles between 90 and 120 degrees (e.g. approximately two hours of observing a target), then the sky variation is only about one magnitude. To address this, we first computed the expected SNR of our Alluxa OD4 filter over a range of sky brightness values (see the left panel of Fig.~\ref{fig:skysens}). We then ran simulations covering the full range of resulting SNR values that we had calculated. The plotted data (right panel, Fig.~\ref{fig:skysens}) demonstrates how the system is would perform over a wide range of SNR values and, by extension, sky brightnesses. As expected, increasing the sky background degrades the performance of the AO system. Additionally, we find that an SNR of 12 appears to be the lower limit of this system, which corresponds to a sky which is three times brighter than our measured value (i.e. over one magnitude brighter). The lower limit comes from the point at which the simulations are too unstable to finish a full set of 20 simulation runs.

There are a few other features worth highlighting in Fig.~\ref{fig:skysens}. First, the point with the highest SNR in Fig.~\ref{fig:skysens} corresponds to the night time sky performance (i.e. $B=0$) so it represents the optimal performance of this system. Second, that the performance levels off with increasing SNR, lending credence to the idea that we have little to gain from using the narrowest achievable bandpass.  For the configuration described in Sec.~\ref{sec:simu} we find that the performance of the Alluxa OD4 filter is robust for sky brightness changes as large as at least one magnitude.  We note that an allowance for sky surface brightness variations should be accommodated in the design of the system and that some characterization of the sky brightness at the telescope site would be useful input to the design.

As mentioned in Sec.~\ref{subsec:apptocots}, the filter profile was measured to be wider than that which the manufacturer provided for us. While the simulations used the filter profile provided to us by the vendor and not the one we measured, our conclusions remain unchanged. The reason for this lies in our sensitivity to sky brightness variations. Calculating the expected SNR of the measured filter profile yields a value of $16$, which still falls within our operable range as demonstrated in Fig.~\ref{fig:skysens}. The increase in filter bandpass is roughly the equivalent of the sky increasing in brightness by a factor of $2$. This does reduce the flexibility of our system's configuration with respect to variations that lead to a brighter sky; however, with an improved bandpass for the interference filter we can regain this flexibility.

\subsection{Optical Layout Considerations}
\label{disc:truncation}
There are also optical design constraints that need to be considered. First, since the sky is such a dominant background source, a field stop at the entrance of the WFS is required to keep the sky background low in each subaperture.  This is especially true for a Shack-Hartmann WFS where each subaperture has the potential of seeing the sky from all adjacent subapertures. Second, the variation of the central wavelength of the filter as a function of angle of incidence sets a field size limitation. For MIRAO the filter needs to pass a field of view of about 4-6 arcseconds to accommodate the elongation of the laser spot as seen by subapertures near the edge of the pupil.  For TMT with a laser launched from behind the telescope secondary mirror (i.e.~on-axis) the elongated spots in the subapertures near the edge of the pupil are expected to be 3-4 arcseconds in length. Assuming the filter is placed at a pupil position in the wavefront sensor, this angular size on the sky is magnified by the pupil magnification at the filter location. For a 30-meter telescope and a 30mm pupil size, this is a factor of 1000.  The subaperture image would be attenuated by 85\% at $\pm 2$ arcseconds.  One can create a larger pupil size and use a larger interference filter but this increases the size and cost of the instrument. Alternatively, one can increase the filter bandpass and compensate for the increased sky background by increasing the size of the subapertures (reducing the overall AO performance).  Finally, one could accept some attenuation of the LGS spot at the edges of the field.  We have not studied the impact of this on the performance of the system and will leave it to a future study.   However, we note that most of the flux will be contained within the center $\sim 3-4$  arcseconds whereas the attenuation is in the tails of the LGS image.  Each of these considerations has their own respective trade-offs which will need to be considered when designing a sodium LGS AO system to be used during the day.

\section{Conclusion}
\label{sec:conc}
Our study indicates that a sodium LGS based AO system can work in daytime conditions with an off-the-shelf narrow-bandpass interference filter and a K-band natural guide star wavefront sensor for tip/tilt/focus sensing. The key filter requirements are a bandpass less than $500\;\mathrm{pm}$, a flat top response with a high (greater than $90\%$) throughput at the central wavelength of the laser guide star return, and an out-of-band suppression of at least OD 4.  The size of the WFS subapertures and the brightness of the LGS needs to be large enough to overcome the still higher sky background.  Simulations of a system like MIRAO for TMT show that a system with 34x34 subapertures, 5 LGSs, and a single on-axis NGS for tip/tilt/focus sensing can deliver Strehl ratios greater than 80\% throughout the thermal infrared range during the day when there is a NGS with $K$ magnitude greater than $13$.  The limiting magnitude of the system is determined by the natural guide star used for tip/tilt/focus correction.

Fine tuning of the filter specifications, the order of the AO system, and optimization of the system control parameters can potentially yield higher Strehl ratio values and fainter limiting magnitudes.  However, this optimization needs to be done in the context of knowledge of the variations of the sky brightness and an operational plan for the daytime observing.  These will be explored in a future study.

This work also has implications for existing LGS AO systems. The ability to operate a sodium LGS in daytime conditions opens up the early twilight hours for system calibrations. This frees up time during the night normally spent setting up and calibrating the system. For some systems this can amount to 20-30 minutes of additional science time per night\cite{wizinowich}.  In addition, with this observations even at near-infrared wavelengths could be extended.  

\subsection*{Disclosures}
The authors have no financial interests in this paper and no potential conflicts of interest to disclose.

\acknowledgments 

We would like to thank the Research Corporation of the University of Hawaii for funding this research through the RCUH Project Development Initiatives Fund. We would also like to thank Mike Connelly, Bobby Bus and the IRTF Day Crew for their help with broadband measurements of the daytime sky with NASA IRTF. Additionally, we would like to thank Lianqi Wang for his help in setting up the MAOS simulations. Finally, we would like to thank Christoph Baranec for his valuable feedback on a draft version of this paper.

\bibliography{report} 
\bibliographystyle{spiejour}   

\vspace{2ex}\noindent\textbf{Ryan Dungee} is a graduate student at the University of Hawaii. He received his BA and MS degrees in Physics from the University of Pennsylvania in 2015 and 2016 respectively. He is working on a PhD in Astronomy currently. His current research interests are ground layer adaptive optics,  astrochemistry, and young stellar objects.

\vspace{1ex}\noindent\textbf{Mark Chun} is an Associate Specialist at the Institute for Astronomy at the University of Hawaii-Manoa.  He received his Ph D. degree in Astronomy and Astrophysics from The University of Chicago in 1997.  He has worked on the deployment of five adaptive optics systems on telescopes around the world.  His current research interests include extending adaptive optics corrections to very wide fields of view and novel wavefront sensing approaches.

\vspace{1ex}\noindent\textbf{Yutaka Hayano} is an Associate Professor at National Astronomical Observatory of Japan. He received his Ph D. degree in Astronomy from The University of Tokyo in 1995. He has worked on the laser guide star adaptive optics system for Subaru Telescope, and ground-to-satellite laser communication. He is currently working for the first generation instrument, IRIS, for TMT. His current research interests are expanding the applications and technologies of adaptive optics.

\listoffigures
\listoftables

\end{spacing}
\end{document}